\documentstyle[epsf,cite,12pt]{article}
\begin{document}
\newcommand{\ome}{\omega_{\rm rot}}
\newcommand{\sumA}{\sum_{i=1}^A}
\newcommand{\sul}{${}^{32}$S}
\newcommand{\boldr}{\mbox{\boldmath$r$}}
\newcommand{\boldj}{\mbox{\boldmath$j$}}
\newcommand{\boldrho}{\mbox{\boldmath$\rho$}}

\title{
High-spin yrast structure of ~${}^{32}$S~ suggested by
symmetry-unrestricted, cranked Hartree-Fock calculations
}

\author{Masayuki Yamagami and Kenichi Matsuyanagi\\
\it Department of Physics, Graduate School of Science,\\
Kyoto University,
\it  Kitashirakawa, Kyoto 606-8502, Japan}
\date{19 August 1999}
\maketitle

\begin{abstract}

High-spin yrast structure of~~\sul~is investigated by means of the
cranked Skyrme-Hartree-Fock method in the three-dimensional Cartesian-mesh
representation without imposing restrictions on spatial symmetries. 
The result suggests that
1)~a crossover from the superdeformed to the hyperdeformed-like
configurations takes place on the yrast line at angular momentum
$I \simeq 24$, which corresponds to the ``band termination'' point
in the cranked harmonic-oscillator model, and 
2)~non-axial octupole deformations of the $Y_{31}$ type 
play an important role in the yrast states 
in the range $5\leq I \leq 13$. \\



\end{abstract}

\vspace{2cm}
\newpage
\section{Introduction}

Since the discovery of the superdeformed(SD) rotational band in
${}^{152}$Dy, about two hundreds SD bands have been found in various 
mass (A=60, 80, 130, 150, 190) regions
\cite{nol88,abe90,jan91,bak95,bak97,dob98}.
It turned out that every regions of superdeformation have their own
characteristics and offer a number of interesting questions; 
investigations of them have been significantly enlarging and 
deepening our understanding of nuclear structure.
Yet, the doubly magic SD band in \sul, 
which has been expected quite a long time~\cite{she72,lea75,rag78,ben81}
remains unexplored, and will become a great challenge in the coming
years~\cite{dob98}.
Exploration of the SD band in \sul~will certainly give a strong
impact toward understanding the possible connection
between the SD structure and 
the molecular resonance phenomena associated with the ${}^{16}$O
+${}^{16}$O configurations~(see, e.g.~\cite{abe80,gre95}~for reviews).
More generally speaking, the nucleus \sul~seems to be situated in a key
position in the investigation of possible relationships 
(such as discussed in~\cite{naz92,fre95,fre97})
between the SD states systematically observed in heavy nuclei and
the cluster structures widely observed in light nuclei~
(see, e.g.~\cite{fuj80}~for a review). 
Thus, excited states in \sul ~have been theoretically studied by 
Nilsson-Strutinsky approaches \cite{she72,lea75,rag78,ben81},
selfconsistent mean-field approaches \cite{gir83, flo84},
spherical shell-models \cite{bre97,kan98}, and
cluster-structure and molecular-resonance points of view 
\cite{zin75,bau80,sch82,zha94,cse98}.

The aim of this paper is to study the high-spin yrast structure of \sul~
from the point of view of exploring exotic shapes in nuclear high-spin
states by means of the cranked Hartree-Fock (HF) method 
with the use of the Skyrme forces~\cite{bei75,bar82},
which is called ``the cranked SHF method''.
One of the recent advances in nuclear structure theory is that
it has become possible to carry out the HF
calculation in the three-dimensional (3D) Cartesian-mesh
representation \cite{dav80,bon85,taj96,taj98}.
This approach has been extended \cite{flo84,flo82,bon87}
to a rotating frame by introducing the cranking term 
and applied to the high-spin yrast states of \sul ~in Ref.~\cite{flo84}
with the use of the BKN interaction \cite{bon76}.
In these cranked HF calculations, however, 
parity and signature symmetries are assumed
for the intrinsic wave functions in order to simplify the calculation. 
We refer an excellent review by \AA berg, Flocard and Nazarewicz~\cite{abe90}
for an overview of studies on nuclear shapes 
in terms of various kinds of mean-field theory, 
especially other than the cranked SHF approach.

Recently, we constructed a new computer code 
for the cranked SHF calculation based on the 
3D Cartesian-mesh representation, which
provides a powerful tool for exploring exotic shapes (breaking both
axial and reflection symmetries in the intrinsic states) at high spin
in unstable nuclei as well as in stable nuclei.
As a first application of this new code, we investigated
the high-spin yrast structure of \sul, and found~\cite{yam98} that
1) a drastic structure change may occur  
above angular momentum $I \simeq 24$ in the yrast line, and
2) non-axial octupole deformations of the $Y_{31}$ type
arise in the yrast line in the range $5\leq I \leq 13$. 
The present paper is intended to give a more detailed account of this work.
Quite recently, Molique, Dobaczewski and Dudek~\cite{mol99}
investigated several SD configurations in \sul~ 
(not restricted to the yrast states) as well as
in neighboring odd-$A$ nuclei by means of the cranked SHF method
with the SLy4 force~\cite{cha98} in the harmonic oscillator basis. 
On the other hand, they did not discuss the yrast states 
above $I \simeq 24$ as well as non-axial octupole deformations, 
which are the major subjects of this paper. 

After a brief account of the cranked SHF calculational method in Section~2, 
an overview of the obtained yrast line for  \sul ~is given in Section~3.
In Section~4, we discuss properties of the high-spin limit of the SD
band, paying special attention to a band-crossing phenomenon associated 
with the level crossing with the rotation-aligned $[440]\frac 12$ level.
The result of the cranked SHF calculation is compared in Section~5 
with that of the cranked harmonic oscillator (CHO) model calculation.
In Section~6, effects of the rotation-induced,
time-odd components in the selfconsistent mean field on the properties 
of the SD band are briefly discussed. 
In Section~7, we discuss about the $Y_{31}$ deformed solutions of the
cranked SHF equations, which constitute
the yrast line in the range $5\leq I \leq 13$. 
Although, at the present time, experimental data directly comparable with
our theoretical calculations seem to be unavailable,
we briefly remark in Section~8 on some recent experimental references.
Conclusion is given in Section~9.

\section{Cranked SHF calculation}

The cranked HF equation for a system uniformly rotating about the
$x$-axis is given by 

\begin{equation}
\delta<H - \ome J_x>=0,
\end{equation}

\noindent
where $\omega_{\rm rot}$ and $J_x$ mean the rotational frequency and
the $x$-component of angular momentum,  
and the bracket denotes the expectation value with respect to a
Slater determinantal state.
We solve the cranked HF equation for a Hamiltonian of the Skyrme type 
by means of the imaginary-time evolution technique\cite{dav80}
in the 3D Cartesian-mesh representation. 
We adopt the standard algorithm
\protect\cite{dav80,bon85,taj96,bon87} in the numerical calculation,
but completely remove various restrictions on spatial symmetries.
Namely, we basically use the procedure developed and applied to
the yrast line of $^{24}$Mg by Bonche, Flocard and Heenen~\cite{bon87},
except that the parity and the signature symmetries are not imposed
on the individual wave functions.
In this connection, we mention that a similar HF code 
(with parity projection but without the cranking term) 
was constructed by Takami et al.~\cite{tak96} 
and successfully applied to the description of 
cluster structures in light nuclei, 
$^{8}$Be,  $^{12}$C, $^{16}$O and $^{20}$Ne.
The same code (but without parity projection) was recently used
to explore exotic shapes in proton-rich
$N \simeq Z$ nuclei in the $^{80}$Zr region~\cite{tak98,mat98}, and
tetrahedral and triangular shapes are suggested
to appear near the ground states of some nuclei in this region.
In Refs.~\cite{bon87,tak98,mat98}, the pairing correlations were taken 
into account in the BCS approximation.
In the present calculation, we neglect the pairing, 
since they are not expected to
play an important role at high-spin states in~ \sul.

When we allow for the simultaneous breaking of both reflection and
axial symmetries, it is crucial to accurately
fulfill the center-of-mass condition

\begin{equation}
<\sumA x_i>=<\sumA y_i>=<\sumA z_i>=0,
\end{equation}

\noindent
and the principal-axis condition 

\begin{equation}
<\sumA x_iy_i>=<\sumA y_iz_i>=<\sumA z_ix_i>=0.
\end{equation}

\noindent
For this purpose we examined several techniques~\cite{yam97} and
confirmed that the constrained HF procedure with quadrupole constraints~
\cite{flo73} works well. Thus, we replace the ``Routhian''
$R=H - \ome J_x$ in eq. (1) with

\begin{equation}
R' = R - \sum_{k=1}^3 \mu_k <\sumA (x_k)_i>^2 
       - \sum_{k<k'}^3 \mu_{k,k'} <\sumA (x_kx_{k'})_i>^2.
\end{equation}

\noindent
In numerical calculations, we confirmed that the constraints (2) and (3) are
fulfilled to the order $O(10^{-15})$ 
with values of the parameters $\mu_k \sim O(10^2)$ and 
$\mu_{k,k'} \sim O(1)$. 
We solved these equations inside the
sphere with radius $R$=8~[fm] and mesh size $h$=1~[fm], starting 
with various initial configurations.
The 11-point formula was used as the difference formula
for the Laplacian operator.
As usual, the angular momentum is evaluated as $I\hbar=<J_x>$.

In this paper, we use the standard SIII and SkM$^*$ forces.
With the use of the SIII force~\cite{bei75}, Tajima et al.~\cite{taj96}
carried out a systematic SHF+BCS calculation for
the ground-state quadrupole deformations of nuclei
in a wide area of nuclear chart. They have carefully examined the
possible error due to the use of the mesh size $h$=1[fm]
and found that the deformation energies obtained with this mesh size
are quite accurate.
On the other hand, the SkM$^*$ force~\cite{bar82} was 
designed to accurately describe 
properties at large deformations like fission barriers, 
so that it may be  suited for the
description of superdeformations~\cite{taj98}.
In recent years, several newer versions of the Skyrme forces have been
proposed (see, e.g. Ref.~\cite{rei99}) in order to improve 
isospin properties of the Skyrme forces. Although the major purpose of
them is to better describing neutron-rich unstable nuclei,  
it will also be interesting to employ such versions to 
examine the dependence of the results reported in this paper 
on the effective interactions adopted. 
We defer such a more extensive calculation for the future.

\section{Structure of the yrast line}

The calculated yrast line is displayed in Fig.~1, and
angular momenta and deformations of the yrast states are drawn  
as functions of rotational frequency in Figs. 2 and 3.
In these and succeeding figures, 
the calculation were done in step of $\Delta\ome=0.2$ MeV/$\hbar$,
and the calculated points (indicated by symbols) are smoothly
interpolated by lines.
The quadrupole deformation parameters $\beta_2$  and $\gamma$  are defined as
\begin{eqnarray}
\beta_2\cos\gamma                                         
&=&\frac {4\pi}{5}~<\sumA r_i^2 Y_{20}(\theta_i,\phi_i)> 
/<\sumA \boldr_i^2> , \\                                     
\beta_2\sin\gamma                                        
&=&-\frac {4\pi}{5}~<\frac 1{\sqrt2}
\sumA r_i^2 (Y_{22}(\theta_i,\phi_i)+Y_{22}^*(\theta_i,\phi_i))>
/<\sumA\boldr_i^2> . 
\end{eqnarray}

It is seen from Figs.~1-3 that the results of calculation 
with the SIII and SkM$^*$ forces are quite similar: For both cases,
the expected SD band becomes the yrast for $I \geq 14$,
and it exhibits a singular behavior at about $I \simeq 24$.
As we shall discuss in the next section, this is due to
a band crossing associated with the rotation-aligned $[440]\frac 12$ level, 
and we call the yrast states
above $I \simeq 24$ ``hyperdeformed (HD)-like
configuration'' in order to distinguish them from the SD configuration.
This configuration becomes unstable against fission for $I \geq 34$.
In addition to the SD and HD-like configurations mentioned above, 
we found that the yrast states with $5\leq I \leq 13$ 
possess an appreciable amount of 
non-axial octupole deformation of the $Y_{31}$ type,
so that we call, for convenience, this region of the yrast line 
``$Y_{31}$ band'', although, as discussed in Section 7,
some cautions are necessary in using this terminology.\\

Thus, the calculated yrast line can be roughly divided into the
following four regions:\\
1)~$I \leq 4$,~~weakly prolate region,\\ 
2)~$ 5 \leq I \leq 13$,~~$Y_{31}$ deformed region,\\
3)~$ 14 \leq I \leq 24$,~~SD region,\\
4)~$ 26 \leq I \leq 32$,~~HD-like region.\\

Below we first discuss the properties of the high-spin limit
of the SD band, and later about the $Y_{31}$ band. 
The lowest-spin region will be touched upon in Section~7 briefly.

\section{High-spin limit of the SD band}

As we saw in Figs.~1-3, the solutions of the cranked SHF equations 
corresponding to the yrast SD configuration are 
obtained from $I=0$ to about $I=22$.

Figure 4 shows the potential energy function for the SD state at $I=0$, 
evaluated by means of the constrained HF procedure~\cite{flo73}
with the quadratic constraint on the mass-quadrupole moment.
We see that the excitation energy of the SD state
at $I=0$ is about 12 MeV. 

A particularly interesting point is the behavior of the SD band
in the high-spin limit: 
It is clearly seen in Figs.~2 and 3 that a jump occurs both in 
the angular momentum $I$ and the quadrupole deformation $\beta_2$  
at $\ome\simeq 2.9$ MeV/$\hbar$.
At this point, $I$ jumps from about 22 to 26, and $\beta_2$
suddenly increases from about 0.6 to 0.7. 
Such a discontinuity is well known~\cite{ham85} 
to occur in the description of the band crossing phenomena 
within a standard framework of the cranked mean-field approach.
The point is more clearly seen in Fig.~5 as a singular behavior of 
the dynamical moment of inertia
${\cal J}^{(2)}=dI/d\ome$ near the band crossing point. 
(Other properties of ${\cal J}^{(2)}$ will be discussed in the next section.)

Figure 6 displays the
shape evolution of the SD band as a function of angular momentum
in the $(\beta_2,\gamma)$ plane:
With increasing angular momentum, small triaxial deformations
gradually set in and at $I \simeq 24$ 
the shape exhibits a striking ``back-bending''
toward larger prolate deformations.
Evidently, this is due to the band crossing mentioned above.
Such a singular behavior of the SD band can be noticed also in the
previous cranked HF calculation with the BKN force~\cite{flo84}.
In Fig.~6 we also plot the $I=$24 and 26 states, which are missing in 
Figs.~1-3, by smoothly extrapolating the $I-\ome$ and
$(\beta_2,\gamma)-I$ curves for the SD and the HD-like configurations,
respectively 
(see, Ref.~\cite{ham85} for the treatment of the band-crossing region).

The microscopic origin of this singular behavior may be understood when we
examine the single-particle energy diagram in the rotating frame
(routhians) presented in Fig. 7.  We see that
the rotation-aligned level associated with the $[440]\frac 12$ orbit
comes down in energy with increasing $\ome$ and crosses
the Fermi level at $\ome\simeq2.9$ MeV$\hbar$ which corresponds to
$I \simeq 24$. 
Thus, the yrast states above $I \simeq 24$ are characterized by
the occupation of the $[440]\frac 12$ level by a single proton and a neutron. 
According to the deformed harmonic-oscillator model, 
N=Z=18 is a magic number associated with
the HD shell structure with axis ratio 3:1,
in which the $[440]\frac 12$ level is occupied by two protons and two neutrons. 
In order to distinguish the yrast states with $I\geq 26$   
from the SD states below $I\simeq24$ and keeping in mind
a connection to the HD configuration, 
we call them ``the HD-like configuration'',
although the magnitude of the quadrupole deformation $\beta_2$ 
obtained in the SHF calculation is in fact comparable to that of 
the SD shape rather than the HD shape.

Let us note that if we regard the SD configuration 
as to correspond to the $j$-$j$-coupling shell model $4p$-$12h$
configuration
$\pi[(f_{7/2})^2(sd)^{-6}]\otimes\nu[(f_{7/2})^2(sd)^{-6}]$
(relative to ${}^{40}$Ca) in the spherical limit, 
the maximum angular momentum that can be
generated by aligning the single-particle angular momenta 
toward the direction of
the rotation axis is $24\hbar$, and thus ``the SD band termination''
might be expected at this angular momentum.
Interestingly, our calculation indicates that a crossover to the
HD-like configuration takes place just at this region of the yrast line.

\section{Comparison with the CHO model}

The behavior at the high-spin limit of the SD band obtained in the
SHF calculation possesses
some similarities with that expected from the CHO model.
This model has been frequently used~\cite{BM,val56,rip75,zel75,
tro79,tro81}
as a simplified model of rotating mean fields.
With obvious notations, the single-particle Hamiltonian of this model 
is written as 

\begin{equation}
h'= \sum_{k=1}^3 \hbar\omega_k(c_k^{\dag} c_k + \frac12) 
- \omega_{\rm rot} l_1,
\end{equation}
where 
\begin{equation}
c_k^{\dag} = \sqrt{m\omega_k/2\hbar}(x_k-ip_k/m\omega_k),
\end{equation}
with $(x_1,x_2,x_3)$ indicating $(x,y,z)$, etc.

The orbital angular momentum operator $l_1$ consists of two parts:
\begin{equation}
l_1=l_1^{(\Delta N=0)}+l_1^{(\Delta N=2)}
\end{equation}
with
\begin{eqnarray}
l_1^{(\Delta N=0)} &=& i\hbar\frac{\omega_2 + \omega_3}
{2\sqrt{\omega_2\omega_3}}(c_3^{\dag} c_2 -c_2^{\dag} c_3), \\
l_1^{(\Delta N=2)} &=& i\hbar\frac{\omega_3-\omega_2}
{2\sqrt{\omega_2\omega_3}}(c_2^{\dag} c_3^{\dag} -c_3c_2). 
\end{eqnarray}

For a given value of $\ome$ or $I\hbar=<\sumA (l_1)_i>$, 
one can determine the oscillator frequencies $(\omega_1 ,\omega_2,
\omega_3)$ such that the selfconsistency condition between the 
density and the potential,

\begin{equation}
\omega_1^2<\sumA (x_1^2)_i>=\omega_2^2<\sumA (x_2^2)_i>=
\omega_3^2<\sumA (x_3^2)_i> ,
\end{equation}
\noindent

\noindent
is fulfilled under a volume conservation condition~\cite{tro81}.
Here, the brackets denote expectation values with respect to 
Slater determinantal states composed of single-particle eigenmodes of $h'$.
\noindent

Let us denote the total number of quanta in
each of the three directions $(k=1,2,3)$ at $\ome=0$ as

\begin{equation}
\Sigma_k=<\sumA(c_k^{\dag}c_k +\frac 12)_i>,
\end{equation}

\noindent
and let us continuously follow the configuration 
specified by the set of values $(\Sigma_1,\Sigma_2,\Sigma_3)$ 
which are defined at $\ome \neq 0$
as the number of quanta associated with the normal modes of the
CHO Hamiltonian $h'$.
In terms of $\Sigma_k$, the selfconsistency condition  at $\ome=0$
is written as

\begin{equation}
\omega_1\Sigma_1=\omega_2\Sigma_2=\omega_3\Sigma_3.
\end{equation}

If the $\Delta N=2$  part of the angular momentum operator $l_1$ 
is neglected, it is well known that
there exists a maximum
angular momentum $I_c=\Sigma_3-\Sigma_2$ 
for a given configuration $(\Sigma_1,\Sigma_2,\Sigma_3)$,
where the shape is oblate and the symmetry axis coincides with the rotation
axis~\cite{BM}.
This shape evolution is caused by the effect of the $\Delta N=0$ part 
of the cranking term,
which tends to align the angular momentum of
individual particles toward the rotation axis of the system
(rotation alignment effect due to the Coriolis force).
In the case of the doubly closed shell configuration for the SD magic
number $N=Z=16$ (including the spin-degeneracy factor 2), 
corresponding to the SD band in \sul,
$(\Sigma_1,\Sigma_2,\Sigma_3)=(24,24,48)$ taking into account
protons and neutrons.
We would thus expect the ``SD band termination'' at
the maximum angular momentum $I_c=\Sigma_3-\Sigma_2=24$.
This number coincides with that evaluated in the previous section
in relation to the $j$-$j$ coupling shell-model configurations. 

On the other hand, the $\Delta N=2$ part stretches the system
toward larger deformations, and actual shape evolutions as functions
of angular momentum are determined by the competition and balance between
these two effects. 
Fully taking into account both effects of the cranking term,
Troudet and Arvieu~\cite{tro79, tro81} found that there is a critical
value $\alpha_c$ of $\Sigma_3/\Sigma_2$,

\begin{equation}
\alpha_c=
\frac{\sqrt{27} + \sqrt 2}{\sqrt{27}-\sqrt 2} \simeq 1.75,
\end{equation}

\noindent
such that the configuration $(\Sigma_1,\Sigma_2,\Sigma_3)$
does not (does) reach the oblate limit 
if $\Sigma_3/\Sigma_2$ is greater (less) than $\alpha_c$.
This is because, for large deformations,
the stretching effect of the $\Delta N=2$ term
dominates at high spin 
over the alignment effect of the $\Delta N=0$ term.
In the case of \sul, the SD configuration have
$\Sigma_3/\Sigma_2=\omega_2/\omega_3=2 > \alpha_c$ at $I=0$.
Therefore, the ``oblate limit'' mentioned above will not be reached 
and the shape at the ``band termination'' point will be triaxial.

Figure 8 shows the shape evolution in the $(\beta_2,\gamma)$ plane,
calculated for the SD configuration of \sul~in the CHO model.
Here, the result of calculation for the configuration (22,24,54)
is also presented, as an example of the HD-like configurations. 
We see that, although the triaxiality slowly sets in 
with increasing angular momentum, the shape of the SD states
remains rather far from the
oblate limit and exhibits a striking ``back-bending''
at about $I_c=24$ toward larger prolate deformations for $I \geq I_c$.
Apparently, the behavior near the critical angular momentum $I_c$
for the SD band is quite similar to that of the SHF solutions 
presented in the previous section.
On the other hand, it should be recalled in comparing Fig.~8 with Fig.~6
that the highest spin region of $I=26 \sim 32$ on the yrast line
corresponds to the HD-like configuration in the SHF solution:
While the continuation of the SD configuration (24,24,48)
to the $I>24$ region as well as that of the HD-like configuration
(22,24,54) to the $I<26$ region are presented for the CHO model, 
only the yrast states were obtained and plotted in the SHF calculation.

Figure 9 shows the angular momentum and the dynamical moment of
inertia ${\cal J}^{(2)}$ as functions of the rotational frequency. 
We see that ${\cal J}^{(2)}$ 
gradually decreases until the critical point.
It is interesting to compare this property with that of 
${\cal J}^{(2)}$  for the SD band in the SHF calculation (Fig.~5).
Apparently, they are quite similar. This suggests that 
the gradual decrease with increasing $\ome$ 
of the dynamical moment of inertia for the SD band 
is rooted in the existence of the
critical angular momentum $I_c$ associated with the quantum
SD shell structure.
We feel that a more detailed investigation of the SD states
near the ``band termination'' 
point is a very important and challenging subject 
for a deeper understanding of the
rotational motion of the nucleus as a finite Fermion system,

\section{Effects of time-odd components}

In this section we shortly discuss about the rotation-induced,
time-odd components in the mean field.
The moment of inertia of the SD band is expected to be 
a good physical quantity for identifying the effects of the time-odd
components, since the pairing correlation plays only a minor role there.
Concerning the effect of various time-odd components on the moment
of inertia, we refer some recent works; \cite{ben94} for a
semiclassical description, \cite{fra96} for a rotating nuclear matter,
and \cite{dob95} for SD bands around $^{152}$Dy. 

Table 1 shows individual contributions from various kinds of time-odd term.
It is interesting to note that the contributions from terms containing
the spin-density, $\boldrho(\boldr)$, 
nearly cancel out with each other and, accordingly, the
contribution from the current-density terms, denoted by $B_3+B_4$, 
dominates in the sum.
Such a remarkable cancellation of the spin-density terms was not seen
in the case of $^{152}$Dy~\cite{dob95}, 
and may be specific to~\sul~under consideration.

In Fig.~10 we compare the results of calculation 
with and without the time-odd components. 
It is seen that the time-odd components increase
the angular momentum for a given value of $\ome$. Accordingly,
the dynamical moment of inertia ${\cal J}^{(2)}=dI/d\ome$ also increases.
This trend is understood from the consideration of the local Galilean
invariance of the Skyrme force \cite{ben94, dob95}
(for a more general analysis not restricted to
the Skyrme force, see \cite{BM,fra96}):
If the time-odd components is neglected, the local Galilean 
invariance is violated and we obtain the moment of inertia
associated with the effective mass $m^*$. 
By including the time-odd components, however, the local Galilean invariance
is restored and we get the moment of inertia associated with the
nucleon mass $m$.
The calculated result presented in Fig.~10 is consistent with this
expectation, but a more quantitative analysis is not necessarily
easy, because, as seen in Fig.~5, 
the calculated moment of inertia significantly deviates from the
rigid-body value due to the shell effect.

\newpage
\noindent
Table 1\\
Contributions from various terms in the time-odd energy density,  
\begin{eqnarray*}
H_{\rm odd}(\boldr) = 
- B_3 \boldj^2  
- B_4 ( \boldj_n^2+\boldj_p^2 )
+ B_9 ( \boldj\cdot \nabla \times \boldrho 
       +\boldj_n\cdot \nabla \times \boldrho_n
       +\boldj_p\cdot \nabla \times \boldrho_p ) \\
+ B_{10}\boldrho^2
+ B_{11}( \boldrho_n^2 +\boldrho_p^2 )
+ B_{12}\rho^\alpha \boldrho^2
+ B_{13}\rho^\alpha ( \boldrho_n^2+\boldrho_p^2 ),
\end{eqnarray*}

\noindent
to the energy (in MeV), calculated at $\ome=1.0$ MeV$/\hbar$ 
for the SIII and SkM$^*$ forces. 
Here, $(\boldj_n,\boldj_p)$ and $(\boldrho_n,\boldrho_p)$
denote the nucleon currents and the spin densities (for neutrons and
protons), respectively, and
$\boldj=\boldj_n+\boldj_p$ and $\boldrho=\boldrho_n+\boldrho_p$
(see Ref.~\cite{bon87} for their explicit expressions).
In columns designated by coefficients $B_i$, 
values after the spatial integration are listed. 
while the total value $\int d\boldr H_{\rm odd}(\boldr)$
and the sum of contributions from the current terms 
(the first two terms in the r.h.s. of the above equation) are shown
in the columns denoted by ``total'' and ``$B_3+B_4$'', respectively.
For reference, the effective mass $m^*$ in nuclear matter for each
force is also listed.\\

\noindent
\begin{tabular}{l||ccccccc||c|c||c} \hline 
& $B_3$ & $B_4$ & $B_9$ & $B_{10}$ & $B_{11}$ & $B_{12}$ & $B_{13}$ & total & $B_3+B_4$   & $m^{*}/m$ \\
\hline 
SIII      & -1.94 & 0.79 & -0.17 & -0.77  & 0.86  & 0.37  & -0.18 & -1.04   & -1.15 & 0.76 \\
SkM$^*$ & -1.83 & 0.90 & -0.38 & -0.44  & 2.45  & 0.00  & -1.65 & -0.95   & -0.93 & 0.79 \\
\hline
\end{tabular}

\vspace{2cm}

\section{$Y_{31}$ deformation}

As mentioned in Section~3, we found that
the yrast states in the region $5\leq I \leq 13$  
possess a significant amount of non-axial octupole deformations 
of the $Y_{31}$ type. 
It should be emphasized that such an exotic deformation is absent
at $I=0$ but emerges at high spin.
It has become possible to get this kind of solution by using
the new cranked SHF code allowing for the simultaneous breaking of both 
axial and reflection symmetries.

As in \cite{tak98}, we define the octupole deformation parameters 
$\alpha_{3m}$ as
\begin{eqnarray}
\alpha_{3m}  &=& (4\pi/3AR^3)<\sumA(r^3X_{3m})_i>,~~~~~
(m=-3,\cdot\cdot\cdot,3)  
\end{eqnarray}
\noindent
with $R=1.2A^{1/3}$ fm. Here $X_{3m}$ is a real basis of the spherical
harmonics,
\begin{eqnarray}
X_{30}=Y_{30},~~~~~~~~                                          \nonumber
X_{3\mid m\mid} &=&\frac 1{\sqrt2}(Y_{3-\mid m\mid}+Y_{3-\mid m\mid}^*) ,\\ 
X_{3-\mid m\mid}&=&\frac {-i}{\sqrt2}(Y_{3\mid m\mid}-Y_{3\mid m\mid}^*),  
\end{eqnarray}
\noindent
where the quantization axis is chosen as the largest and 
smallest principal inertia axes for prolate and oblate solutions, respectively.
The yrast solutions in the region $5\leq I \leq 13$ have
$\alpha_{31}\ne 0$ but $\alpha_{3m}=0$ for $m \ne 1$.
(See \cite{fra99} for a general discussion on this kind of deformation
and its consequence on rotational spectra.)
Figure 11 shows the calculated values of the $Y_{31}$ deformation 
as a function of $\omega_{\rm rot}$.
We see that the $\alpha_{31}$ value quickly rises 
when $\omega_{\rm rot}$ exceeds 1 MeV/$\hbar$.

The microscopic origin of the growth of the non-axial octupole
deformation $\alpha_{31}$ may be understood when we
examine the single-particle energy diagram in the rotating frame
(routhians) presented in Fig. 12. 
We note that a strong coupling and a quasi-level crossing
between the rotation-aligned $[330]\frac 12$ orbit and 
the $[211]\frac12$ orbit take place near the Fermi surface 
in the region $1.0\leq\ome\leq 2.2$ MeV/$\hbar$.
The matrix element of the $r^3Y_{31}$ operator between the two
single-particle states is large, since they satisfy the selection rule 
for the asymptotic quantum numbers $(\Delta\Lambda=1, \Delta n_z=2)$.
This strong coupling is responsible for
the $\alpha_{31}$ deformation
appearing in  this region of the yrast line.

Figure 13 shows the potential energy function with respect to the
$\alpha_{31}$ direction, calculated by means of the constrained HF
procedure. Note the scale of the ordinate. Although we obtain a 
clear minimum at a finite value of $\alpha_{31}$, the potential is
rather shallow in this direction, so that the
amplitude of the quantum-mechanical zero-point vibrational motion
might be larger than the equilibrium deformation. 
If this is the case, a treatment of dynamics going beyond the mean-field 
approximation is required in order to investigate the consequence 
of the $\alpha_{31}$ deformation on the quantum spectra in the yrast region
under consideration. This is beyond the scope of the present paper.

It may be desirable to extend the potential energy curve in Fig. 13
to the $\alpha_{31}=0$ limit. It turned out, however, difficult to do so,
because many level-crossings take place with decreasing $\alpha_{31}$.
[If we extrapolate to this limit assuming parabolic dependence on 
$\alpha_{31}$, we obtain about 2 MeV as a very crude estimate of the
energy gain due to the $\alpha_{31}$ deformation.] 
By the same reason, it is also difficult to follow the continuation
of the $Y_{31}$ band to the higher spin region as soon as it departs
from the yrast line.

\section{Some remarks on experimental data}

Although rich experimental data are available for excited states
of \sul, the high-spin yrast region in which we are interested 
is rather poorly known at the present time.
Accordingly, we discuss experimental references only briefly.

For low-spin states with $I \leq 7$, detailed spectroscopic data 
are available up to excitation energy 11.76 MeV~\cite{bre97,kan98}.
These excited states are shown to be well described by the spherical
shell model calculations~\cite{bre97,kan98}. In these works,
some negative-parity states were interpreted as octupole-quadrupole 
phonon multiplets. As a matter of fact, we need to go beyond the simple
mean field theory in order to discuss such spectroscopic
data in the low-spin region.

Highly excited states have been studied by various nuclear reactions
as well as $^{16}$O-$^{16}$O~ scatterings.
Investigating the $^{16}$O($^{20}$Ne$,\alpha)^{32}$S$(\alpha)^{28}$Si
(g.s.) reaction, Morita et al~\cite{mor85} suggested possible band structures 
of the quasi-molecular configuration of $^{16}$O$+^{16}$O and of some
parity-doublet-like structures with angular momenta $5^-, 6^+, (7^-),
(8^+)$ at the 12-15 MeV region. 
Recently, Curtis et al~\cite{cur96} investigated the region with 
$I$=10-16 and the excitation energy 32-38 MeV by means of the
$^{12}$C$(^{24}$Mg$,^{16}$O$^{16}$O$)^{4}$He reaction, and
suggested an existence of highly deformed states in this region.
It is tempting to compare these experimental data with our theoretical 
calculations. The experimentally explored regions are, however,
about 10 MeV above the theoretical yrast line.
Therefore, a more detailed spectroscopic study is needed
in order to associate these data with the yrast structure.

\section{Conclusions}

We have investigated the high-spin yrast structure of ~\sul~by means of the
cranked SHF method in the 3D Cartesian-mesh representation
without imposing restrictions on spatial symmetries, 
and suggested that\\
1)~a crossover from the SD to the HD-like
configurations takes place on the yrast line at angular momentum
$I \simeq 24$, which corresponds to the ``band termination'' point
in the CHO  model, and\\
2)~non-axial octupole deformations of the $Y_{31}$ type 
play an important role in the yrast states 
in the range $5\leq I \leq 13$. 

In conclusion, we would like to stress again that the calculated
yrast line for $I$= 14-20 lies about 10 MeV below the observed
molecular resonance region associated 
with the $^{16}$O-$^{16}$O configurations.
Thus, a yrast $\gamma$-spectroscopy with higher resolving power 
is strongly desired in order to explore the
high-spin region of the yrast line of ~\sul~.

\section*{Acknowledgements}

During the course of this work, we have benefited from discussions
with J. Dobaczewski, P.-H. Heenen, M. Matsuo, W. Nazarewicz, N. Tajima,
S. Takami and K. Yabana.  
We would like to express our hearty thanks to them.
We are grateful to S. Frauendorf for pointing out the importance of
distinguishing the $\alpha_{31}$ and the $\alpha_{3-1}$ deformations
and sending us his article \cite{fra99} before publication.
This work was supported by the Grant-in-Aid  for Scientific
Research (No. 10640264) from the Japan Society for the Promotion of Science.


\newpage
\begin{figure}
\epsfxsize=16cm
\centerline{\epsffile{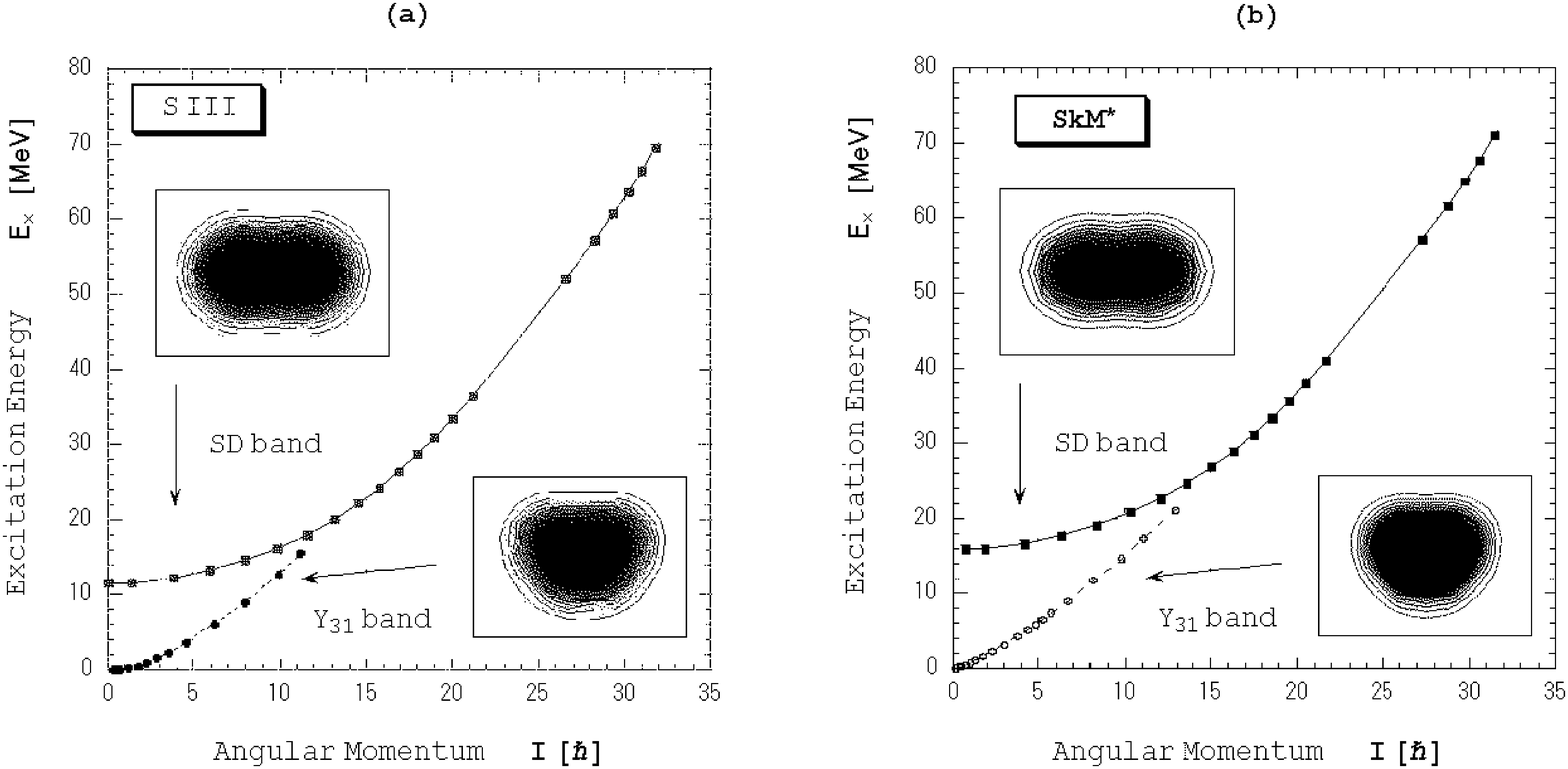}}
\caption{
(a) Excitation energy vs. angular-momentum plot 
for the yrast structure 
of~\sul, calculated with the SIII force.
Density distributions on the plane {\it perpendicular} to the 
rotation axis are shown, as insets, for 
the SD band (solid line) and the $Y_{31}$ band (dashed line).
The calculation was done in step of $\Delta\ome=0.2$ MeV/$\hbar$,
and the calculated points (indicated by symbols) are smoothly
interpolated by lines.~~
(b) Same as (a), but with the SkM$^*$ force.
}
\vspace*{10pt}
\label{fig1}
\end{figure}

\begin{figure}
\epsfxsize=8cm
\centerline{\epsffile{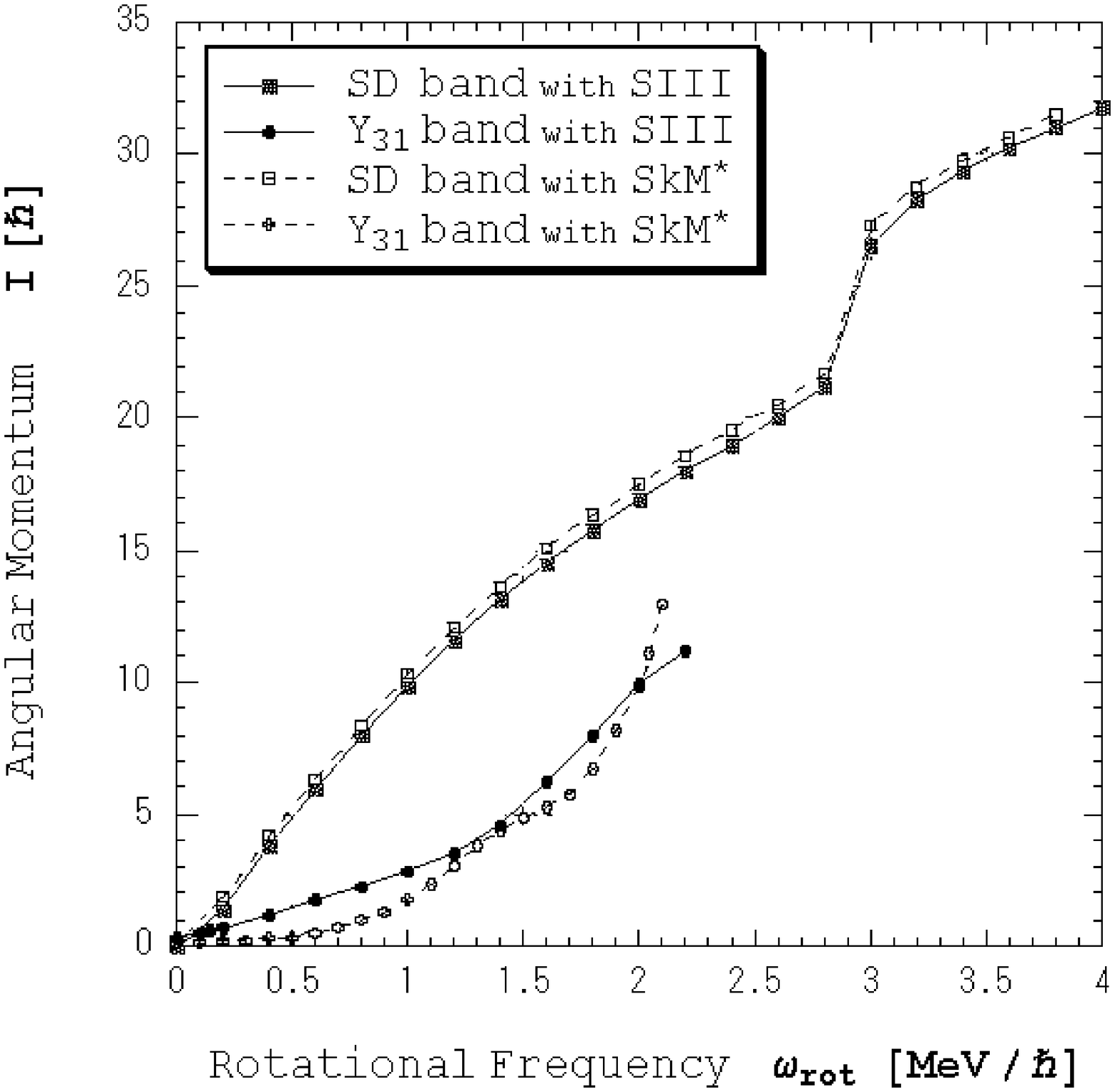}}
\caption{
Angular momentum $I$ plotted as a function of rotational
frequency $\ome$ for the SD band and the $Y_{31}$ band in~\sul.
Results calculated with the SIII and SkM$^*$ forces are shown
by solid and dashed lines, respectively.
}
\vspace*{10pt}
\label{fig2}
\end{figure}

\begin{figure}
\epsfxsize=8cm
\centerline{\epsffile{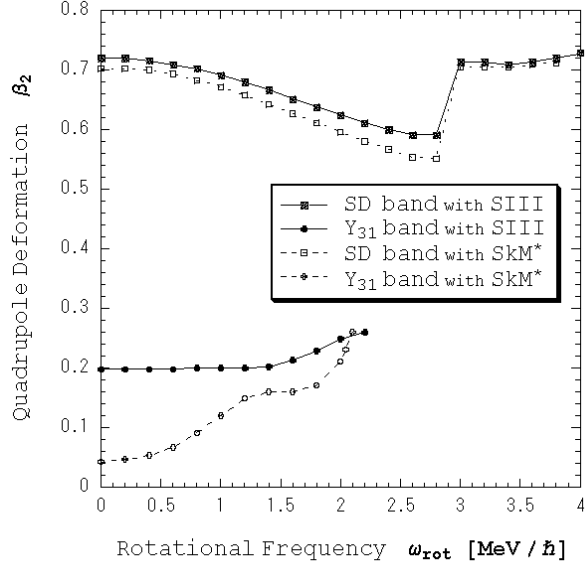}}
\caption{
Quadrupole deformation $\beta_2$ plotted as a function of rotational
frequency $\omega_{rot}$ for the SD band and 
the $Y_{31}$ band in~\sul. Results calculated with the SIII 
and SkM$^*$ forces are shown by solid and dashed lines, respectively. 
}
\vspace*{10pt}
\label{fig3}
\end{figure}

\begin{figure}
\epsfxsize=8cm
\centerline{\epsffile{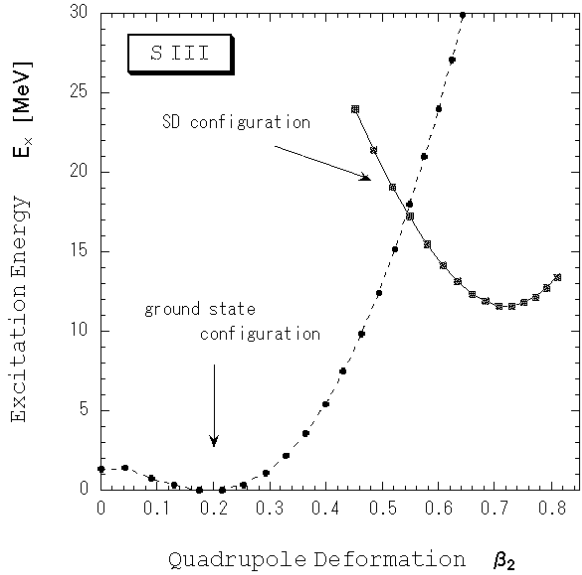}}
\caption{
Potential energy function at $I=0$ for the SD configuration 
(solid line) relative to that for
the ground state configuration (dashed line) in~\sul,
calculated with the SIII force.
}
\vspace*{10pt}
\label{fig4}
\end{figure}

\begin{figure}
\epsfxsize=8cm
\centerline{\epsffile{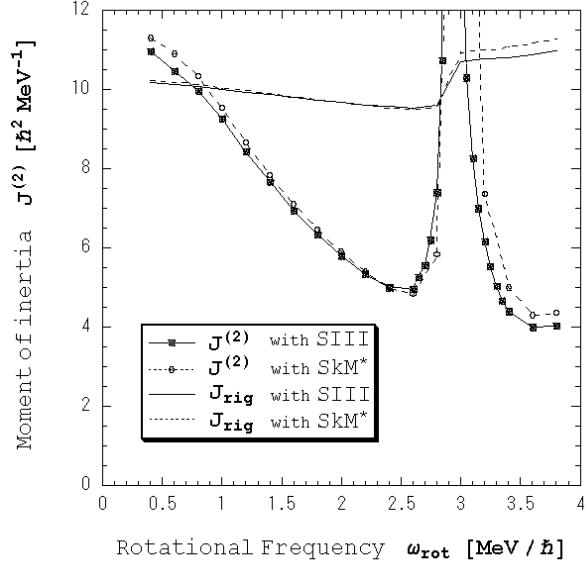}}
\caption{
Dynamical moment of inertia ${\cal J}^{(2)}= dI/d\ome$ 
plotted as a function of $\ome$ for the SD band in~\sul.
Results calculated with the SIII and SkM$^*$ forces are shown
by solid and dashed lines, respectively. 
For reference, the rigid moments of inertia
${\cal J}_{\rm rig}=m\int \rho({\boldr})(y^2+z^2)d{\boldr}$
with the calculated density $\rho({\boldr})$ are also indicated.
}
\vspace*{10pt}
\label{fig5}
\end{figure}

\begin{figure}
\epsfxsize=8cm
\centerline{\epsffile{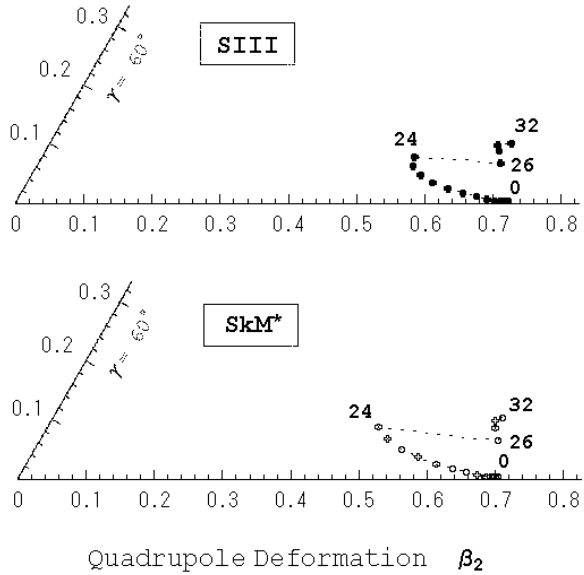}}
\caption{
Shape evolution as a function of angular momentum, 
plotted ~in the $(\beta_2,\gamma)$ plane
for the SD and HD-like configurations in~\sul. 
Results calculated with the SIII and SkM$^*$ forces are separately
shown. 
}
\vspace*{10pt}
\label{fig6}
\end{figure}

\begin{figure}
\epsfxsize=8cm
\centerline{\epsffile{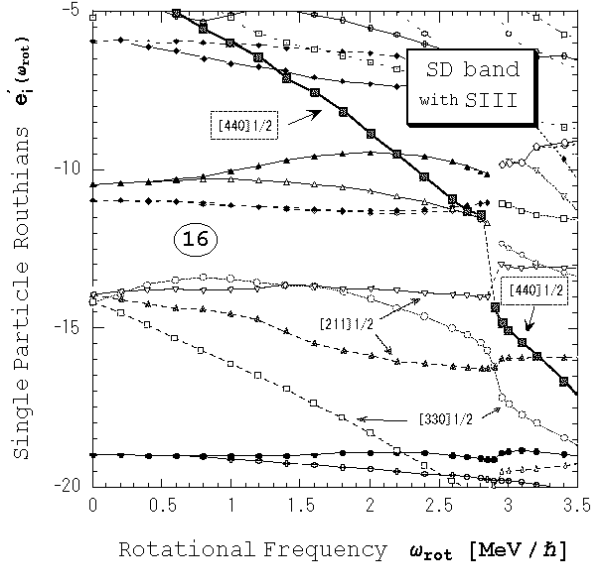}}
\caption{
Single-particle energy diagram (for neutrons) in the rotating frame
for the SD band in~\sul, plotted as a function of $\ome$.
The SIII force is used.
Note that the structure of the yrast configuration drastically
changes at $\ome \simeq 2.9$ MeV/$\hbar$, so that the diagram is
discontinuous about this point, although levels characterized by
the same asymptotic quantum numbers are linked by lines.  
}
\vspace*{10pt}
\label{fig7}
\end{figure}

\begin{figure}
\epsfxsize=8cm
\centerline{\epsffile{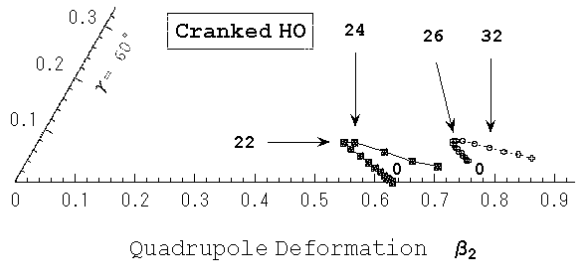}}
\caption{
Shape evolutions as functions of angular momentum
in the $(\beta_2,\gamma)$ plane,
plotted with filled and open symbols, respectively,
for the SD configuration
$(\Sigma_1,\Sigma_2,\Sigma_3)=(24,24,48)$ 
and the HD-like configuration (22,24,54) in the CHO model.
}
\vspace*{10pt}
\label{fig8}
\end{figure}

\begin{figure}
\epsfxsize=16cm
\centerline{\epsffile{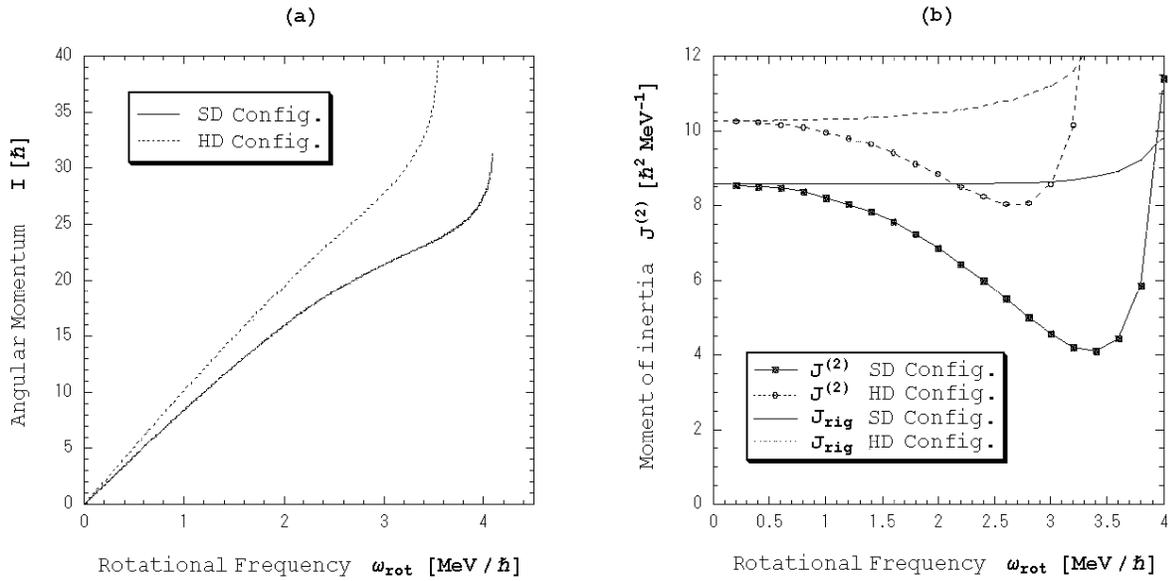}}
\vspace{1cm}
\caption{
(a) Plot of angular momentum vs. rotational frequency 
in the CHO model.
Solid line is used for the SD configuration 
$(\Sigma_1,\Sigma_2,\Sigma_3)=(24,24,48)$,
while dashed line for the HD-like configuration (22,24,54).~~
(b) Same as (a), but for dynamical moment of inertia ${\cal J}^{(2)}
=dI/d\ome$. 
For reference, rigid moments of inertia
${\cal J}_{\rm rig}=m<\sumA(y^2+z^2)_i>$
for these configurations are also indicated.
}
\vspace*{10pt}
\label{fig9}
\end{figure}

\begin{figure}
\epsfxsize=16cm
\centerline{\epsffile{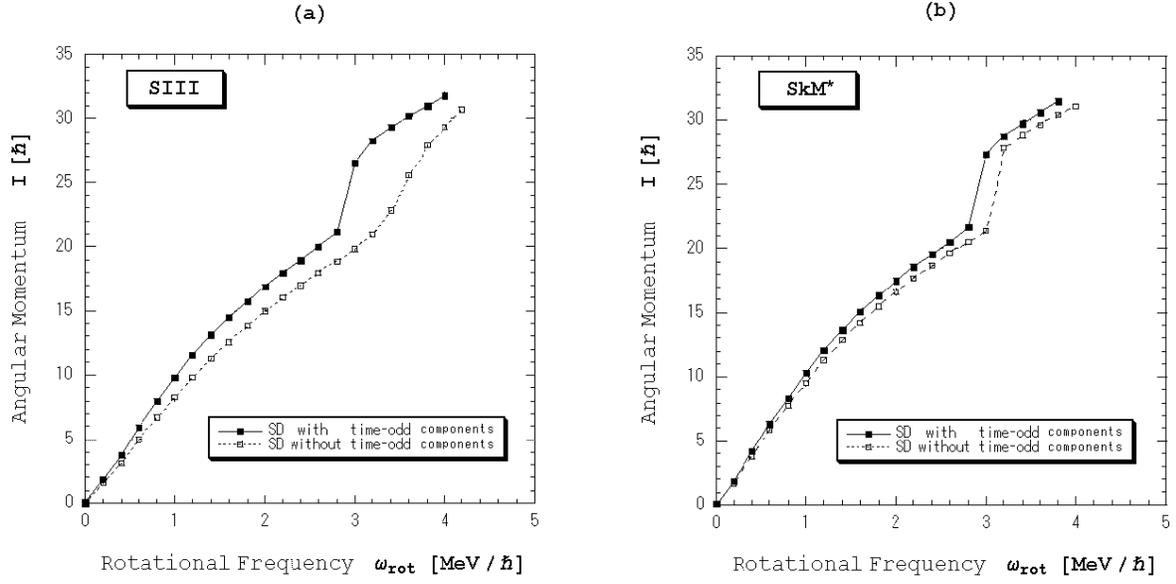}}
\caption{
(a) Angular momentum $I$ plotted as a function of $\ome$
for the SD band in~\sul. 
Solid line with filled squares (dashed line with open squares)
indicates the result with(without) the time-odd components.
The SIII force is used.~~ 
(b) Same as (a), but with the SkM$^*$ force.
}
\vspace*{10pt}
\label{fig10}
\end{figure}

\begin{figure}
\epsfxsize=8cm
\centerline{\epsffile{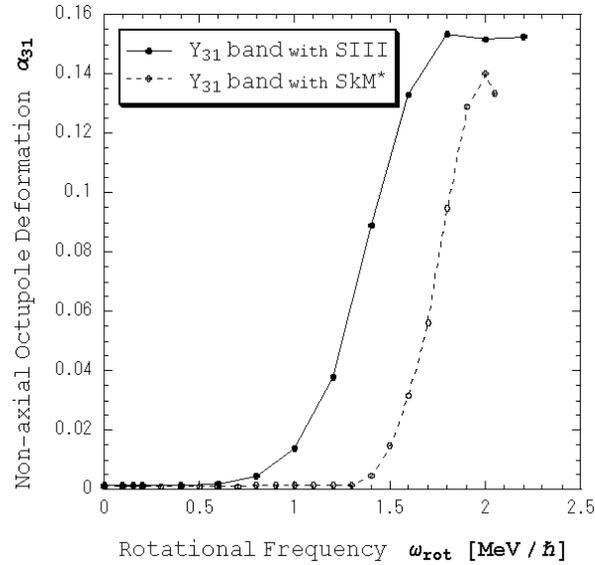}}
\caption{
Non-axial octupole deformation $\alpha_{31}$ 
plotted as a function of $\ome$ for the $Y_{31}$ band in~\sul. 
Results calculated with the SIII and SkM$^*$ forces are shown
by solid and dashed lines, respectively.
}
\vspace*{10pt}
\label{fig11}
\end{figure}

\begin{figure}
\epsfxsize=8cm
\centerline{\epsffile{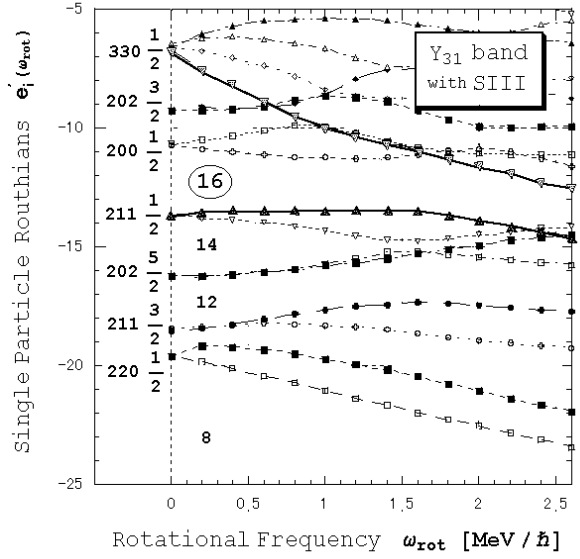}}
\caption{
Single-particle energy diagram (for neutrons) in the rotating frame
for the $Y_{31}$ band in~\sul, plotted as a function of $\ome$.
The SIII force is used.
}
\vspace*{10pt}
\label{fig12}
\end{figure}

\begin{figure}
\epsfxsize=8cm
\centerline{\epsffile{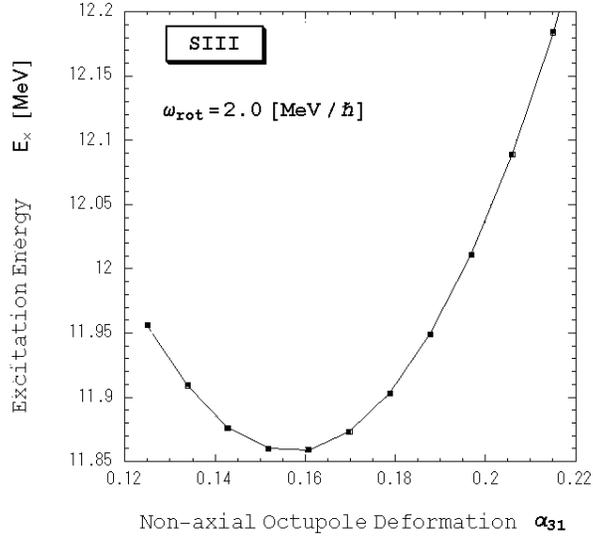}}
\caption{
Potential energy function for the $Y_{31}$ band in~\sul~ at 
$\ome=2.0$ MeV/$\hbar$, calculated by means of the constrained HF
procedure with the SIII force. Note the scale of the ordinate.
}
\vspace*{10pt}
\label{fig13}
\end{figure}

\end{document}